\documentclass[aps,prl,twocolumn,superscriptaddress,showpacs]{revtex4-1}

\usepackage{amssymb,amsfonts,amsmath}
\usepackage{units} 
\usepackage{braket}
\usepackage{relsize} %use \mathsmaller to reduce size of math in captions and methods section

\usepackage{epsfig}
\usepackage{graphicx}

\makeatletter
\DeclareRobustCommand{\genericinterval}[2]{%
  \@ifstar{\genericinterval@star{#1}{#2}}{\genericinterval@nostar{#1}{#2}}}
\newcommand{\genericinterval@star}[4]{\mathopen{}\mathclose{\left#1#3,#4\right#2}}
\newcommand{\genericinterval@nostar}[4]{\mathopen{#1}#3,#4\mathclose{#2}}

\makeatother

\begin{document}
\newcommand{\vq}{\mathbf{q}}
\newcommand{\vp}{\mathbf{p}}
\newcommand{\vP}{\mathbf{P}}
\newcommand{\Wcm}{~Wcm$^{-2}\,$}
\newcommand{\Ea}{E_\mathrm{a}}
\newcommand{\Up}{U_\mathrm{p}}
\newcommand{\Upm}{U_{\mathrm{p},\max}}
\newcommand{\tp}{\tau_\mathrm{t}}
\newcommand{\tr}{\tau_\mathrm{r}}
\newcommand{\Tp}{T_\mathrm{p}}
\newcommand{\tg}{\tau_\mathrm{g}}
\newcommand{\vE}{\mathbf{E}}
\newcommand{\vA}{\mathbf{A}}
\newcommand{\ver}{\mathbf{r}}
\newcommand{\valpha}{\mbox{\boldmath{$\alpha$}}}
\newcommand{\vpi}{\mbox{\boldmath{$\pi$}}}
\newcommand{\Rep}{\mathrm{Re}\,}
\newcommand{\Imp}{\mathrm{Im}\,}
\newcommand{\AL}{A_\mathrm{L}}
\newcommand{\ve}{\hat{\mathbf{e}}}
\newcommand{\Ep}{E_\mathbf{p}}

\newcommand{\etal}{\emph{et al.~}}

\newcommand{\red}[1]{\textcolor{red} {\bf {#1}}}

\title{A Streak Camera for Strong-Field Ionization}

\author{M. K\"ubel}
\email{matthias.kuebel@physik.uni-muenchen.de}

\affiliation{Joint Attosecond Laboratory, National Research Council and University of Ottawa, Ottawa, Ontario, Canada}
\affiliation{Department of Physics, Ludwig-Maximilians-Universit\"at Munich, D-85748 Garching, Germany}

\author{Z. Dube}
\affiliation{Joint Attosecond Laboratory, National Research Council and University of Ottawa, Ottawa, Ontario, Canada}

\author{A. Yu.~Naumov}
\affiliation{Joint Attosecond Laboratory, National Research Council and University of Ottawa, Ottawa, Ontario, Canada}

\author{M. Spanner}
\affiliation{Joint Attosecond Laboratory, National Research Council and University of Ottawa, Ottawa, Ontario, Canada}

\author{G. G. Paulus}
\affiliation{Institute for Optics and Quantum Electronics, Universit\"at Jena, D-07743 Jena, Germany}
\affiliation{Helmholtz Institute Jena, D-07743 Jena, Germany}

\author{M. F. Kling}
\affiliation{Department of Physics, Ludwig-Maximilians-Universit\"at Munich, D-85748 Garching, Germany}
%\affiliation{Max Planck Institute of Quantum Optics, D-85748 Garching, Germany}

\author{D. M. Villeneuve}
\affiliation{Joint Attosecond Laboratory, National Research Council and University of Ottawa, Ottawa, Ontario, Canada}

\author{P. B. Corkum}
\affiliation{Joint Attosecond Laboratory, National Research Council and University of Ottawa, Ottawa, Ontario, Canada}

\author{A. Staudte}
\email{andre.staudte@nrc-cnrc.gc.ca}
\affiliation{Joint Attosecond Laboratory, National Research Council and University of Ottawa, Ottawa, Ontario, Canada}

\date{\today}

%\begin{article}
\begin{abstract}
{Ionization of an atom or molecule by a strong laser field produces sub-optical cycle wave packets whose control has given rise to attosecond science. The final states of the wave packets depend on ionization and deflection by the laser field, which are convoluted in conventional experiments.
Here, we demonstrate a technique enabling efficient electron deflection, separate from the field driving strong-field ionization.
Using a mid-infrared deflection field permits one to distinguish electron wave packets generated at different field maxima of an intense few-cycle visible laser pulse. 
We utilize this capability to trace the scattering of low-energy electrons driven by the mid-infrared field. Our approach represents a general technique for studying and controlling strong-field ionization dynamics on the attosecond time scale.}
\end{abstract}

%\pacs{33.80.Rv, 33.80.Wz, 32.80.Rm, 32.80.Wr}
\maketitle

In the attosecond streak camera an intense optical field is used to temporally resolve single photon ionization caused by an attosecond extreme ultra-violet pulse \cite{Itatani2002,Kienberger2004,Cavalieri2007,Sansone2010,Forg2016,Ossiander2016}. The streaking concept has been applied to characterize electron wave packets and light fields primarily in the extreme ultraviolet, where single photon ionization prevails, e.g. \cite{Goulielmakis2004, Fruehling2009, Schuette2011, Ardana-Lamas2016}. However, an optical field can also time-resolve multiphoton or tunnel ionization. The "attoclock" technique \cite{Eckle2008} exploits the deflection of the photoelectron wavepacket in an intense elliptically polarized near-infrared pulse to address questions regarding time delays \cite{Eckle2008b,Pfeiffer2012} and non-adiabaticity \cite{Landsman2015} in tunnel ionization. 
However, using the same optical frequency for ionization and streaking limits the versatility of this approach.

A relatively weak control field is sufficient to significantly manipulate strong-field interactions. For example, tunnel ionization at every or every second half cycle can be enhanced or suppressed using a third \cite{Watanabe1994} or second \cite{Schumacher1994} harmonic field with parallel polarization, respectively. Orthogonally polarized two-color pulses (e.g.,\cite{Zhang2014,Richter2015}), and elliptically polarized two-color pulses (e.g., \cite{Wu2013PRA,Eckart2016,Mancuso2016}) open other avenues to manipulate strong-field interactions.  %However, since these conventional two-color techniques rely on many-cycle pulses, the tunnel ionization from individual optical cycles cannot be resolved.

Here, we demonstrate STIER (Sub-cycle Tracing of Ionization Enabled by infra-Red), a streak camera that temporally resolves strong-field ionization caused by a linearly polarized few-cycle pulse. We employ STIER to demonstrate the imaging of individual ionization bursts, which occur at the field maxima of a few-cycle laser pulse. This provides insight into the sub-cycle dynamics of strong field ionization. We observe the emergence of an asymmetry in the yield of low-energy electrons associated with re-scattering \cite{Corkum1993} in the ionic potential. Such low-energy rescattering has been linked to low-energy features in photoelectron spectra generated by mid-infrared laser fields \cite{Blaga2009,Wu2012,Wolter2015}, to frustrated tunnel ionization \cite{Nubbemeyer2008}, and stabilization of atoms against ionization in intense fields \cite{Eberly1993,Morales2011,Larimian2016}. The latter leads to the production of highly-excited Rydberg atoms by the intense laser field \cite{Eichmann2009}. With STIER we can trace and control the underlying processes.

STIER samples the photoelectrons produced by a few-cycle laser pulse in the near-visible spectral range, here 735\,nm. The photoelectrons are streaked by a moderately intense, mid-infrared (IR) pulse at 2215\,nm with stable carrier-envelope phase (CEP). The pulse duration of the visible pulse is significantly shorter than the period of the IR pulse, and much shorter than in recent streaking experiments on nanotips using Terahertz pulses \cite{Wimmer2014}. 
The intensities of the light fields are chosen such that ionization only occurs in the presence of the visible pulse. Although the IR pulse does not ionize the target gas, it significantly deflects the generated photoelectrons because the quiver energy of a free electron in a laser field scales with the square of the laser wavelength. Our technique permits the usage of arbitrary polarization states for the two light fields. Here, we choose the polarization of both fields as linear and parallel to each other. Besides control over the deflection of photoelectrons, the parallel polarization also enables control over ionization.

%Condition for streaking: photoionization is confined to a half-cycle of the streaking field.

%According to the principle of the streak camera, the final momentum acquired by each electron wave packet equals the vector potential at the time the wave packet was generated.

\begin{figure}[t]
%\vspace{-0.5}
\centerline{\includegraphics[width=0.5\textwidth]{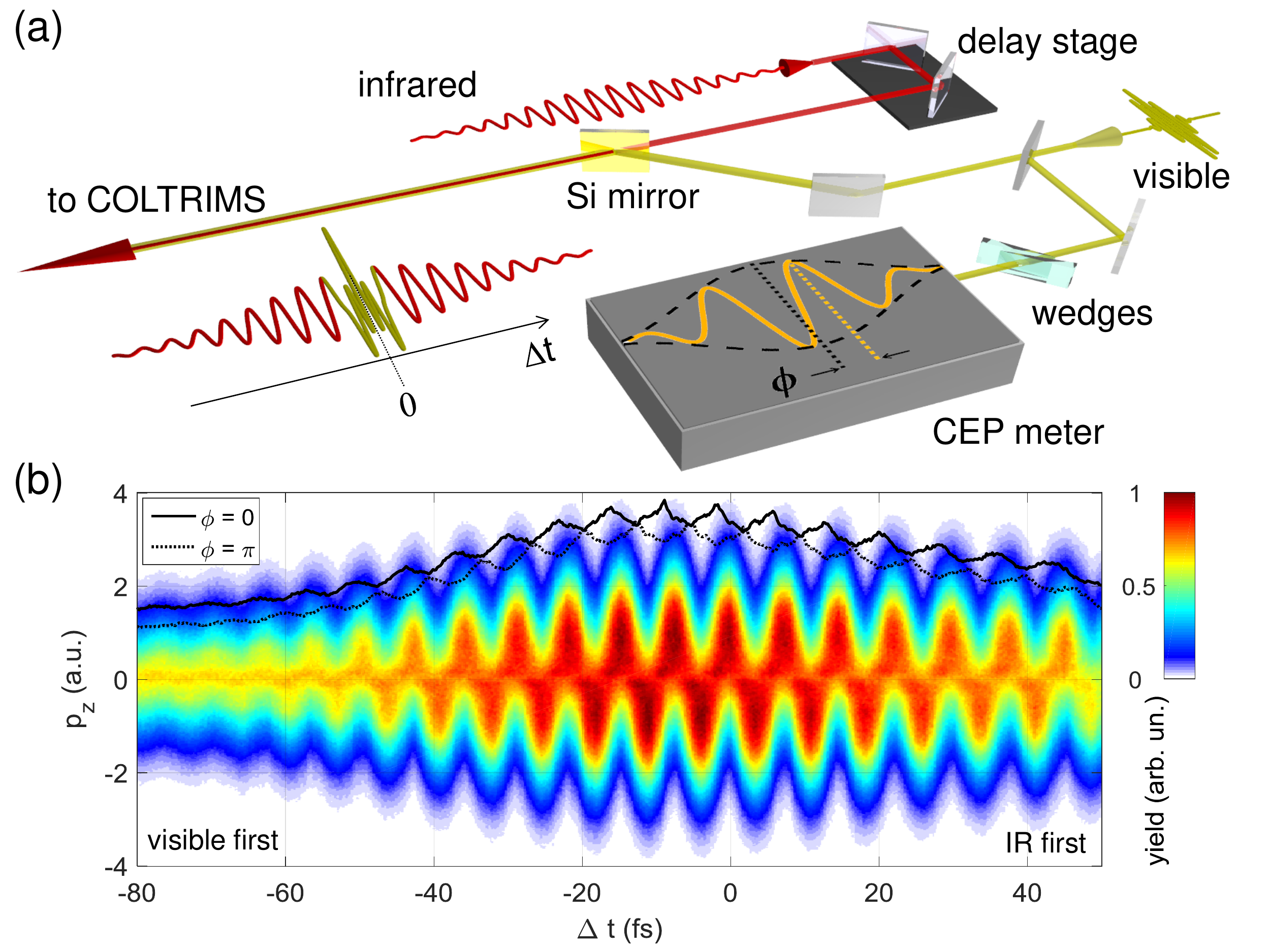}}
%\vspace{-3.5cm}
\caption{(a) Experimental setup for STIER. Few-cycle visible pulses (yellow) with a duration of 5\,fs are split in two parts using a broadband beam splitter. The reflected part (40\%) is sent to the CEP meter. The transmitted part is recombined with a 75-fs IR pulse (red beam) on a silicon mirror at $60^\circ$ angle of incidence. The IR pulse is delayed with respect to the visible pulse using a piezo-driven translation stage. (b) STIER trace recorded in Ne using COLTRIMS, and averaged over CEP. Shown is the ion yield as function of time delay and recoil momentum along the laser polarization. The black lines indicate the delay dependent ion yields for cosine pulses with the field maximum pointing up ($\phi=0$) or down ($\phi=\pi$). The curves are offset for visibility.} 
\label{fig:experiment}
\end{figure}

The experimental setup is sketched in Fig.~\ref{fig:experiment}. The output from a commercial amplified laser system (Coherent Legend Elite Cryo, 1.6\,mJ, 800\,nm, 10\,kHz), is split in two parts to obtain CEP stable IR pulses from an optical parametric amplifier (Light Conversion TOPAS-Prime), and few-cycle visible pulses from an argon filled hollow-core fiber. The visible pulses are phase tagged using a stereographic above threshold ionization phase meter \cite{Paulus2003, Rathje2012}. This yields the CEP of each laser pulse with an unknown but constant offset.

%and VIS, respectively. 
%In the IR part, an optical parametric amplifier (Light Conversion TOPAS-Prime) is pumped with 1.35\,mJ to generate CEP-stable pulses at a wavelength of 2250\,nm. A piezo-driven translation stage is used to delay the IR pulses with respect to the visible pulses. 
%In the VIS part of the beam path, few-cycle pulses are obtained via spectral broadening in an argon-filled hollow-core fiber and adequate pulse compression using chirped multilayer mirrors \cite{Pervak2009} and fused silica wedges. A broadband beam splitter is used to direct 40\% of the available power towards a stereographic above-threshold ionization CEP meter. 
%The linear polarization of the visible pulses can be rotated using a achromatic half-wave plate (HWP). 
%Recombination of the IR and visible pulses occurs on a silicon mirror at $60^\circ$ angle of incidence. 

After recombination, the two-color pulses are sent to a cold target recoil ion momentum spectrometer (COLTRIMS) \cite{Ullrich2003} where they are focused into a neon gas jet.  %In the COLTRIMS, t
The three-dimensional momentum vectors of ions and electrons are recorded with COLTRIMS and correlated with the delay $\Delta t$ between IR and visible pulses, and with the CEP $\phi$ of the visible pulses. %We focus our analysis on the ion recoil momenta along the laser polarization.%
%We note that STIER with parallel polarization can be readily adapted to velocity map imaging and time-of-flight spectroscopy. %and COLTRIMS is therefore not required.

The two-color laser field (see Figs.~\ref{fig:experiment}(a,b)) can be written as 
\begin{equation}
\vec{E} (t) = E_\mathrm{VIS}(t) \vec{e_z} + E_\mathrm{IR}(t) \vec{e_z},
\label{field definition}
\end{equation}
where 
\begin{align}
E_\mathrm{IR}(t+\Delta t)&= \sqrt{I_\mathrm{IR}(t+\Delta t)} \cos(\omega_\mathrm{IR}(t+\Delta t)), \\
E_\mathrm{VIS}(t)&= \sqrt{I_\mathrm{VIS}(t)} \cos(\omega_\mathrm{VIS}t+\phi). 
\label{eq:VIS_IR_pulse}
\end{align}
The intensity envelopes $I_\mathrm{VIS}(t)$, and $I_\mathrm{IR}(t)$ are characterized by a full width at half maximum of 5\,fs for the visible, and 75\,fs for the IR pulse. The frequencies $\omega_\mathrm{VIS/IR}$ correspond to the visible and IR wavelengths of 735\,nm and 2215\,nm, respectively. 

%For parallel polarization between IR and visible pulses geometry,
Experimental results are compared to computational results obtained by solving the one-dimensional (1D) time-dependent Schr\"odinger equation (TDSE) for a 1D soft core Coulomb potential using the Fourier split-operator method. The initial ground state was found by complex time propagation. To limit the computational demand, the IR pulse is approximated as a monochromatic field with a duration of 4.25 cycles and 0 field strength at the beginning of the simulation. The peak intensities are $I_\mathrm{VIS} = \unit[7\times 10^{14}]{Wcm^{-2}}$ and $I_\mathrm{IR} = \unit[3\times 10^{13}]{Wcm^{-2}}$. 
Focal volume averaging for the visible pulse, including integration over the Gouy phase, is taken into account.  Averaging over the Gouy phase of the IR pulse has the same effect as a jitter of the relative time delay between IR and VIS pulses. Based on the experimental data, we estimate the uncertainty in the time delay as $\unit[\pm 0.8]{fs}$.

%The experimental data is further compared to the predictions from a 2D semi-classical model for which the ionization rate at each point in time during the laser pulse is calculated using the rates given in  \cite{Tong2005}. When the ionization rate is non-negligible, an electron is placed at the tunnel exit with initial momentum zero along the laser polarization and a random value in accord with Ref.~\cite{Delone1991} perpendicular to the polarization. The trajectories are propagated until the end of the laser pulse by solving the equation of motion (in atomic units)
%\begin{equation}
%\ddot{\vec{r}}=-\vec{E}(t) - \vec{r} / (|\vec{r}|^2+\alpha)^{3/2}, 
%\end{equation}
%with the soft-core parameter $\alpha=1$. The 2D momentum distribution is generated from all trajectories with a positive final energy. For each delay step and CEP value, $10^6$ trajectories are calculated.

Fig.~\ref{fig:experiment}(b) shows a STIER spectrogram recorded in Ne. %for parallel and perpendicular geometries are 
The momentum distribution along the polarization axis exhibits strong delay-dependent oscillations with a period of 7.4\,fs, corresponding to the optical period of 2220\,nm light.  
The oscillation amplitude reaches a maximum of $\Delta p \approx \unit[1.4]{a.u.}$ at the center of the IR pulse, which relates to an IR intensity of $I_\mathrm{IR} \approx \unit[(3 \pm 1) \times 10^{13}]{Wcm^{-2}}$. The width of the momentum distribution in the absence of the IR field indicates an intensity of $I_\mathrm{VIS} \approx \unit[(7 \pm 2) \times 10^{14}]{Wcm^{-2}}$. The strong oscillations are evidence that ionization is essentially confined to a half-cycle of the IR field. Thus, the IR vector potential is imaged by the delay dependence of the observed momentum distributions.

In order to resolve the discrete ionization events in the few-cycle visible pulse, we now sort the STIER traces by CEP. This fixes the phase of the visible pulse to the phase of the IR field. The black lines in \ref{fig:experiment}(b) show how the IR field modulates the ionization probability for two different CEP values of the visible pulse. Changing the CEP by $\pi$ turns a yield maximum into a yield minimum.

\begin{figure}[t]
%\vspace{-0.5cm}
\centerline{\includegraphics[width=.47\textwidth]{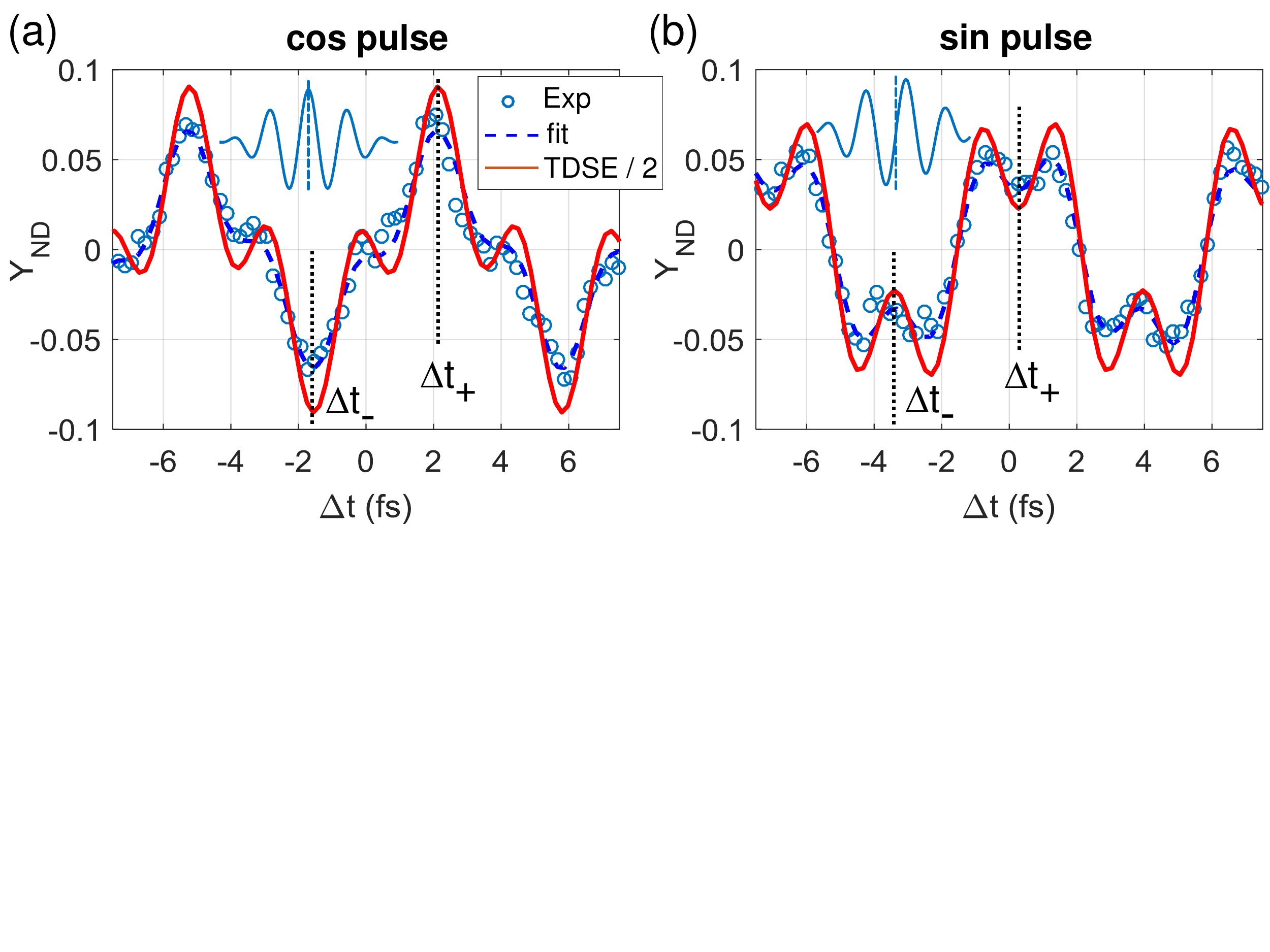}}
\vspace{-3cm}
\caption{Delay dependence of the total ionization yield for (a) cosine and (b) sine visible pulses. %using parallel polarization. 
Shown is the normalized difference of the yields at CEPs $\phi$ and $\phi+\pi$ (equation \ref{eq:ND-yield}), integrated over a range of $\pm \pi/8$ each. Delay values with enhanced (attenuated) yield are indicated by the dotted lines labeled $\Delta t_+$ ($\Delta t_-$). The TDSE result is scaled by a factor of 0.5 to allow for a better comparison of the yield periodicity with the experiment. The dashed blue line represents a sinusoidal fit with two frequencies.}
\label{fig:ND_yields_1d}
\end{figure}

In Fig.~\ref{fig:ND_yields_1d} the IR-induced modulation of the ionization yield is analyzed in detail for two waveforms, corresponding to cosine ($\phi = n \pi$, $n=0, 1$) or sine ($\phi = (2n+1)\pi/2$) pulses. We introduce the normalized difference of the ionization yield $Y(\phi)$ as 
\begin{equation}
Y_\mathrm{ND} = (Y(\phi)-Y(\phi+\pi))/(Y(\phi)+Y(\phi+\pi)).
\label{eq:ND-yield}
\end{equation}
The symmetry operation $\phi \rightarrow \phi+\pi$ corresponds to inversion of the visible field direction at every point in time, while the direction of the IR field is unchanged. Hence, through definition, $Y_\mathrm{ND}$ reveals the influence of the IR field on the ionization yield.

The data exhibit clear oscillations with two different frequencies. This demonstrates that the ionization probabilities at different field maxima are modulated by the IR streaking field. For cosine pulses (Fig.~\ref{fig:ND_yields_1d}(a)) a single maximum and minimum per IR period exists. For sine pulses (Fig.~\ref{fig:ND_yields_1d}(b)), maximum and minimum yields are obtained at two different time delays per IR period. This can be readily understood by considering that the field strength reaches its maximum value only once during a cosine pulse but twice during a sine pulse.
The fast oscillations are not as pronounced in the experimental data, which we attribute to the timing uncertainty of visible and IR pulses. %However, the shapes of the experimental curves clearly indicate the contribution of more than one frequency. 

The measured values for $Y_\mathrm{ND}$ are fitted with (dashed lines Figs \ref{eq:ND-yield}(a,b))
\begin{align}
Y_\mathrm{ND}(\Delta t)= &A_1 \cos{\left(\omega_1 (\Delta t-t_1)\right)}\\
+&A_2 \cos{\left(\omega_2 (\Delta t-t_1) + \phi_2\right)}.
\end{align}
The fitted optical periods are $2\pi / \omega_1 = \unit[2.45 \pm 0.02]{fs}$ and $2\pi / \omega_2 = \unit[7.38 \pm 0.02]{fs}$, corresponding to laser wavelengths 735\,nm, and 2215\,nm, respectively. 
The absolute CEP of the visible pulse is given by $\phi = \phi_2/2$, which yields the unknown constant offset in the CEP measured by the CEP meter. Moreover, the absolute time delay $\Delta t$ (up to a multiple of the IR period) is given by $t_1$. The knowledge of CEP and delay enables the accurate comparison of experimental and calculated data. 

\begin{figure*}[t]
%\vspace{-0.5cm}
\centerline{\includegraphics[width=1\textwidth]{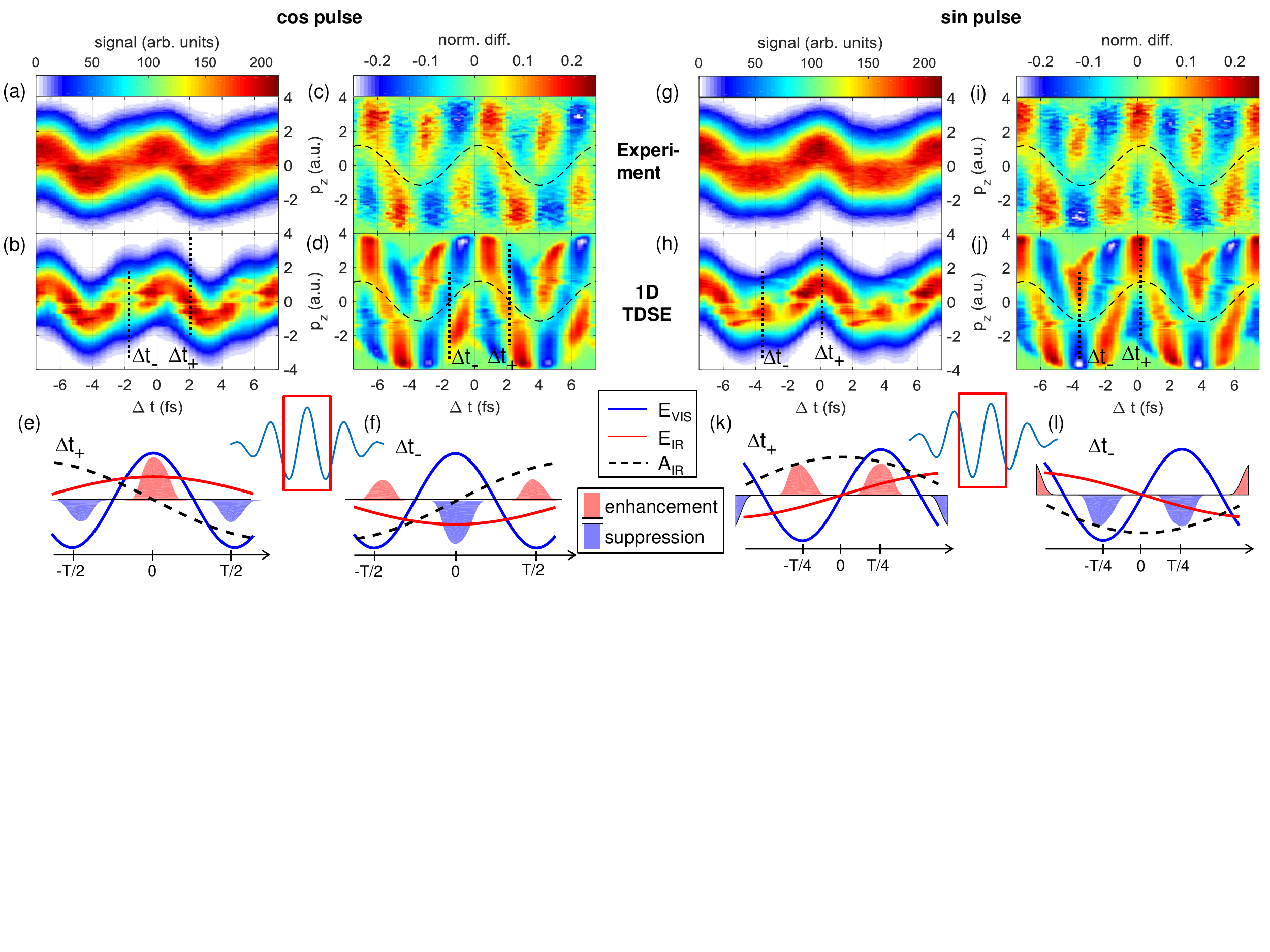}}
\vspace{-5cm}
\caption{Results for cosine($\phi = 0$) pulses (a-d), and sine ($\phi = \pi/2$) pulses (g-l). %using parallel polarizations of VIS and IR pulses. 
The panels in the first (a,b) and third column (g,h) display the recorded and calculated STIER spectrograms for fixed CEP values. The panels in the second (c,d) and fourth column (i,j) show difference spectrograms of the data to their left with respect to the data at $\phi+\pi$. The color scale for the calculated results covers the range [-0.6,0.6]. The dashed black lines indicate the values of the vector potential at the center of the visible pulse at each delay. Delay values with enhanced (suppressed) yield are indicated by vertical dotted lines labeled $\Delta t_+$ ($\Delta t_-$). In panels (e,f,k,l), the most relevant parts of the visible and IR laser fields, as indicated by the red boxes, are illustrated for delay values $\Delta t_+$ and $\Delta t_-$. The red (blue) shaded areas are differential ionization rates \cite{Tong2005}, indicating enhancement (suppression) of the ionization probability at the field maxima of the visible pulse, induced by the IR field.}
\label{fig:1d_stier}
\end{figure*}

The IR field controls the ionization probability at the field maxima of the visible pulse, separated by 1.2\,fs each. During this time, the IR field and its vector potential significantly vary. Therefore, the wave packets generated at different field maxima are deflected to different final momenta. In the STIER traces for fixed CEP values, shown in Fig.~\ref{fig:1d_stier}, we analyze how the IR field controls both ionization and deflection of the generated photoelectrons.

%For simplicity, only the results of the semi-classical calculations are shown here. 
%The CEP and delay dependence of the results obtained by the semi-classical calculations agree qualitatively very well with the ones obtained by solving the TDSE, which are displayed in the Supplementary Material. Thus, the experimental results may be interpreted in terms of the semi-classical model. 
The measured signal for cosine (Fig.~\ref{fig:1d_stier}(a)) and sine (Fig.~\ref{fig:1d_stier}(g)) pulses have distinct shapes. Although some features observed in the calculated signals (Fig.~\ref{fig:1d_stier}(b,h)) are not visible in the experimental data, the characteristics at the yield minima ($\Delta t_-$) and yield maxima ($\Delta t_+$) can be clearly distinguished. %In the case of the cosine pulse, the signal exhibits a slow rising slope at $\Delta t_-$ and a steep falling slope at $\Delta t_+$. In the 
However, the individual ionization bursts throughout the pulse are not visible.%As above, we attribute this to the limited stability of the interferometer.
 
To improve the visibility of the differences in the STIER spectrograms for different CEPs, difference spectrograms are calculated analogously to equation \ref{eq:ND-yield} and displayed in Figs~\ref{fig:1d_stier}(c,d,i,j).
The normalized difference reveals distinct patterns that depend on the CEP and vary on a sub-femtosecond timescale. The experimentally observed patterns agree qualitatively very well with those in the computational results. 

In the following, we show that the difference spectrograms image the IR-induced modulation of the ionization probability at different half-cycles of the visible pulse. As discussed for Fig.~\ref{fig:ND_yields_1d}, the normalized difference reveals the influence of the IR field on the ionization probability. Furthermore, wave packets generated at different field maxima of the few-cycle visible pulse are shifted in momentum by the IR vector potential. In particular, the momentum shift for wave packets created at the center of the visible pulse, i.e., at $t=0$, is given by the IR vector potential $A_\mathrm{IR}(\Delta t)$, which is drawn as a dashed line in Figs \ref{fig:1d_stier}(c,d,i,j). %the IR vector potential maps the ionization time and momentum. 

In Fig.~\ref{fig:1d_stier}(e), the most relevant parts of the visible and IR fields are drawn for a cosine pulse at $\Delta t = \Delta t_+$. In this case, the field maxima of visible and IR coincide and the combined fields lead to an ionization enhancement at $t=0$ (indicated by the red shaded area). Ionization at the field extrema at $t=\pm T/2$ ($T$ being the visible optical period), on the other hand, is suppressed (indicated by the blue shaded area). This is reflected by the corresponding difference spectrograms Fig.~\ref{fig:1d_stier}(c,d), where a maximum is observed at $(\Delta t, p_z)=(\Delta t_+,0)$, and minima are observed for smaller and larger momenta. The situation is reversed at $\Delta t_-$ (see Fig.~\ref{fig:1d_stier}(f)), where the ionization at $t=0$ is suppressed, and ionization at $t=\pm T/2$ is enhanced. This leads to the positive off-center signals around $\Delta t_-$ in Figs \ref{fig:1d_stier}(c,d). Similar analyses can be performed for the sine pulse, as illustrated in Figs \ref{fig:1d_stier}(k,l). For the fields depicted here, the IR vector potential peaks at $t=0$, such that the wave packets generated at $t=\pm T/4$ acquire the same streaking momentum, and cannot be distinguished. 

\begin{figure}[t]
%\vspace{-0.5cm}
\centerline{\includegraphics[width=.5\textwidth]{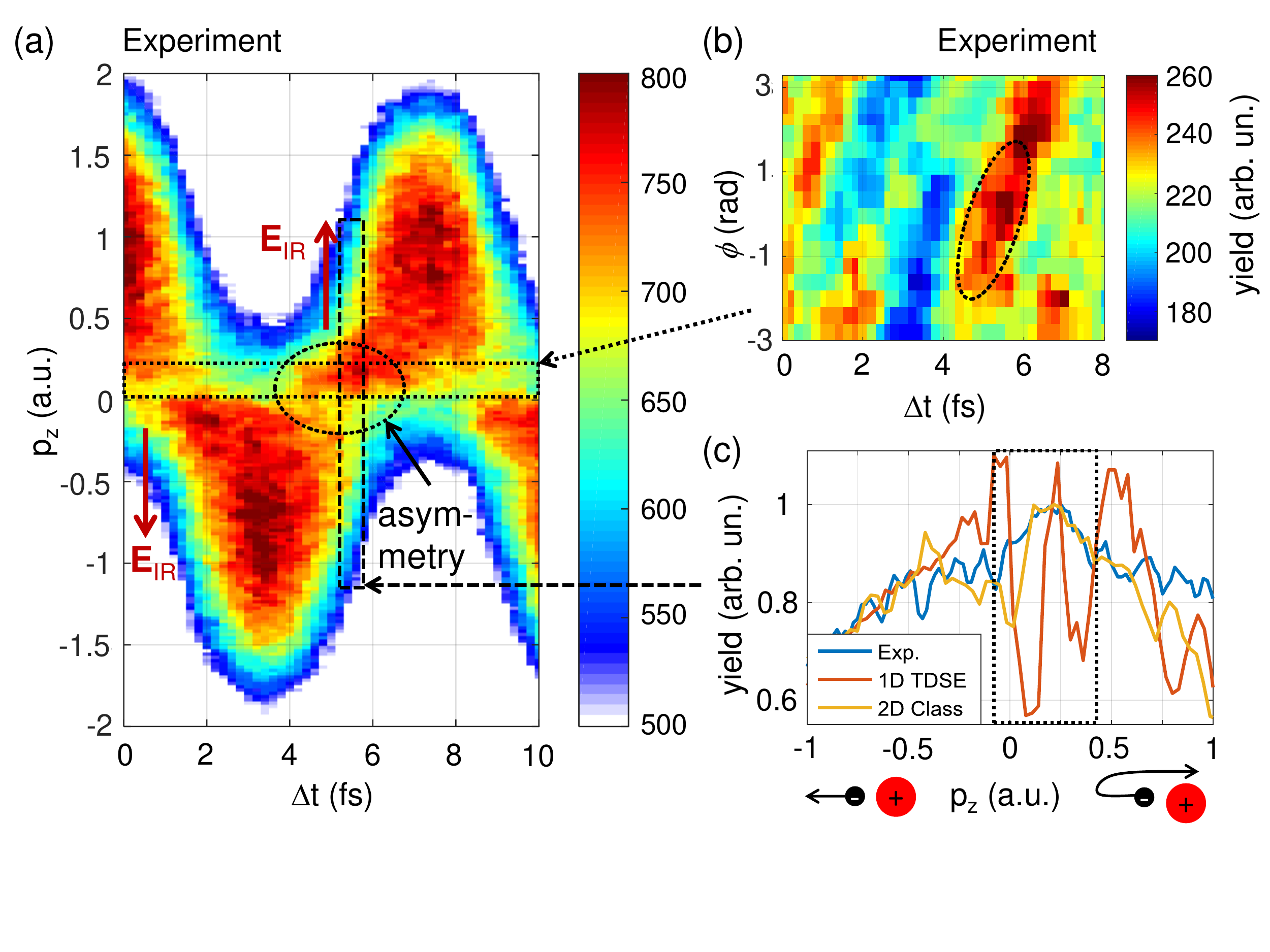}}
\vspace{-1cm}
\caption{Asymmetry feature for up and down streaking. (a) Close-up of the STIER spectrogram measured for Neon, averaged over CEP. The dashed oval indicates the yield enhancement giving rise to an asymmetry between positive and negative momenta. (b) Delay and CEP dependence of the measured yield in the range $0.05<p_z<0.25$, indicated by the dotted box. (c) Momentum spectrum along the laser polarization for a cosine pulse (pointing up) at $\Delta t=\unit[5.6]{fs}$ for experiment, 1D TDSE and 2D classical trajectory Monte-Carlo simulation. The yield enhancement is marked by the dotted box.}  %(c) Same as (b) for the results of the 1D TDSE2D semi-classical simulation. \textbf{Enhance contrast more. Show lineout instead of CEP dependence of theory.}}
\label{fig:hockeystick}
\end{figure}

%In the analysis above, streaking of strong-field ionization was reasonably well described using the so-called simple man's model. 
In Fig.~\ref{fig:1d_stier}, we have shown that STIER permits us to separate the electron wave packets generated at different half-cycle maxima of the visible pulse. The best separation occurs when the visible pulse is centered around a field maximum of the IR pulse, i.e.~when the signal in the STIER trace is centered around $p_z=0$. In this case, the observed momentum distribution directly reflects the ionization dynamics during a single half-cycle. In particular, one half of the momentum spectrum along the laser polarization corresponds to direct electrons that drift in the same direction as they tunnel out of the atom. The other half corresponds to rescattered electrons that turn around after tunneling.

In Fig.~\ref{fig:hockeystick}, we concentrate on the low-energy electrons in the STIER trace, where an asymmetry feature appears for certain delay values. The feature covers positive (negative) momenta for delay values at which the IR field points up (down), see Fig.~\ref{fig:hockeystick}(a). The low-energy yield also depends on the CEP of the visible pulse (Fig.~\ref{fig:hockeystick}(b)). For CEP values at which the strongest half-cycle of the visible field points into the same direction as the IR field, the asymmetry is maximized. Hence, the tunneling direction is opposite to the final momentum of the enhanced low-energy electrons, implying that these electrons are due to a rescattering effect. %Due to their low energy, the involved electrons can be expected to be sensitive to the ionic potential. 

As shown in Fig.~\ref{fig:hockeystick} (c), the 1D TDSE result agrees well with the experimental data for negative momenta, corresponding to direct electron emission. For rescattered electrons, which acquire positive final momenta, however, experiment and 1D TDSE strongly deviate. Most evidently, the pronounced modulations near the maximum measured at $p_z=\unit[0.15]{a.u.}$ are not observed experimentally. 
However, recollision effects are not correctly captured in one dimension. 

A 2D classical trajectory Monte-Carlo simulation yields much better agreement with the experimental data around $p_z=\unit[0.15]{a.u.}$. Based on this model, we can attribute the asymmetry feature to multiple rescattering in the IR field after ionization at a suitable time. While the IR field is present, electrons oscillate in the vicinity of the ion. After the IR field is turned off, the electrons drift into the direction opposite to the one they tunneled out of the atom. Certain trajectories do not acquire significant kinetic energy at the end of the pulse and remain bound in the ionic potential, leading to the minimum at $p_z=\unit[0]{a.u.}$. Rescattering in the visible field leads to larger momenta.

The present low-energy asymmetry feature and excitation phenomena in strong fields \cite{Nubbemeyer2008,Eichmann2009} result from the rescattering of low-energy electrons. Only electrons born with zero energy at times $t_r$, where $A_\mathrm{IR}(t_r) \approx 0$, can revisit the core. Using STIER, the time delay controls whether or not such electrons are generated and thereby controls low-energy recollisions and the resulting phenomena.

We have shown that STIER enables sub-femtosecond time resolution utilizing strong-field ionization by linearly polarized few-cycle pulses. In particular, one may tag a detected electron by the time of the half-cycle when it was emitted. %\emph{(introduce this idea earlier?)}
The possibility to distinguish adjacent half-cycles where the electric field points into opposite direction makes the technique ideal to probe ionization from anisotropic targets such as oriented molecules. 

The introduced approach is general as it does not require specific laser wavelengths and permits different polarization geometries. The case of parallel polarization discussed here enables control over both ionization and deflection. In the case of perpendicular polarization of the two pulses, the influence of both fields is separated into perpendicular directions. The electron momentum distribution in the polarization plane will provide access to the temporal momentum distribution of strong-field produced wave packets. 
Similar to the attoclock technique, the photoelectron momentum distribution can be spread out over a section of a torus when the IR field is (near-)circularly polarized. STIER allows for a new pump-probe scheme when a second visible few-cycle pulse driving ionization is added. This scheme will be useful for the tracing of electronic wave packets as the IR streaking can be exploited to separate the signals from the two few-cycle pulses in momentum space. %Usage of an even longer wavelength for the IR field would be beneficial to reduce overlap from the contributions of different half-cycles of the visible pulse. 

STIER can be applied to a variety of strong-field phenomena, such as double ionization and channel-resolved ionization in molecules. Moreover, the asymmetric fields used in STIER can be utilized to coherently controlling photochemical reactions. %Finally, the effect of the IR streaking field on high harmonic generation remains to be explored.

\begin{acknowledgments}
We thank D.~Crane and B.~Avery for technical support. We are grateful to A.~Czasch for his help with the data acquisition. This project has received funding from the EU’s Horizon2020 research and innovation programme under the Marie Sklodowska-Curie Grant Agreement No. 657544. Financial support from the National Science and Engineering Research Council Discovery Grant No. 419092-2013-RGPIN is gratefully acknowledged. We acknowledge support by the German Research Foundation through the cluster of excellence "Munich Center for Advanced Photonics".
\end{acknowledgments}

\bibliography{stier1_v1}

\end{document}